\documentclass[11pt,a4paper]{article}

 \usepackage[english]{babel}
 \usepackage[T2A]{fontenc}
 \usepackage[cp1251]{inputenc}
 \usepackage{amsmath}
\usepackage{graphicx}
\usepackage{amssymb}
\usepackage{color}
\usepackage{amsfonts}
\usepackage{wrapfig}
\usepackage{caption}

\newcommand{\non}{\nonumber \\}

\newcommand{\be}{\begin{equation}}
\newcommand{\ee}{\end{equation}}

\newcommand{\lp}{\left (}
\newcommand{\rp}{\right )}

\newcommand{\vk}{\mathbf{k}}
\newcommand{\vl}{\mathbf{l}}
\newcommand{\cB}{{\cal B}}
\newcommand{\rhok}{\rho_{\mathbf{k}}}
\newcommand{\rhomk}{\rho_{-\mathbf{k}}}

\begin{document}

\binoppenalty=10000
\relpenalty=10000

\begin{center}
\textbf{\Large{PHASE BEHAVIOR OF A CELL FLUID MODEL WITH MODIFIED MORSE POTENTIAL}}
\end{center}

\vspace{0.3cm}

\begin{center}
M.P.~Kozlovskii and O.A.~Dobush\footnote{e-mail: dobush@icmp.lviv.ua}
\end{center}

\begin{center}
Institute for Condensed Matter Physics of the National Academy of Sciences of
Ukraine \\ 1, Svientsitskii Str., 79011 Lviv, Ukraine
\end{center}

 \vspace{0.5cm}

\small The present manuscript gives a theoretical description of the first-order phase transition in a cell fluid model with a modified Morse potential and additional repulsive interaction.  In the framework of the grand canonical ensemble, the equation of state of the system in terms of chemical potential-temperature and terms of density-temperature is calculated for a wide range of density and temperature. The behaviour of the chemical potential as a function of temperature and density is investigated. The maximum and minimum admissible values of the chemical potential, which approach each other with decreasing temperature, are exhibited. The existence of a liquid-gas phase transition in a limited temperature range below the critical $ T_c $ is established.

\vspace{0.5cm}

PACs: 51.30.+i, 64.60.fd


\normalsize

\section{Introduction} \label{introduction}

Construction of the equation of state is a topical problem of studying the phase behavior of a system of interacting particles. Nowadays, several such equations are, to a great extent, phenomenological in nature and well describe the properties of real substances. There are few approaches in which such equations were obtained at the level of microscopic description. Among them are methods based on investigating the behavior of the virial equation of state~\cite{ref11,ref12}. The latter describes the behavior of systems at the thermodynamic boundary. Nevertheless, the consideration of higher viral coefficients causes problems in displaying the liquid branch of the phase diagram~\cite{ref13}. Another well-known approach to describe the phase behavior of fluids is the theory of integral equations. This method is mostly a numerical procedure for structural analysis of a system based on the calculation of paired correlation functions. One of its implementations is the self-consistent  Ornstein-Zernike approximation (SCOZA)~\cite{ref14, ref15}.

Previously \cite{ref1,ref2,ref3} we proposed a cell fluid model applied to describe a first-order phase transition using different types of interaction potentials.  Particularly in \cite{ref1} the grand partition function is calculated, and the equation of state of the system with the Curie-Weiss potential is obtained. Without using any approximations, we established that such system possesses a cascade of first-order phase transitions describing a sequence of phases with increasing density. The emergence of multiple phase transitions, rather than a single one, was associated with particles being point-like. The explanation of the appearance of high-density phases in this model is such that a term forbidding the excessive concentration of particles in a cell is absent in the interaction potential. Usually, while describing fluid systems, the hard-spheres (or soft-spheres) potential is used, limiting the maximum density \cite{ref4,ref5}. In the present paper, a supplemental term of soft repulsion is included in the interaction potential, and afterwards, an appropriate equation of state is calculated. For this purpose, a modified Morse potential is used, which, along with the attractive and repelling components, contains a part describing the additional soft wall repulsion. The inclusion of such a term in the interaction potential enable to specify a distinct reference system, which makes it possible to calculate the Jacobian of transition from density variables to collective variables. The set of collective variables is natural for describing collective effects, including phase transitions.

The novelty of this work consists in the extension of the previously obtained results \cite{ref6} over a wide temperature range (including the temperature at which the liquid-gas phase transition region ends). This is made possible by the use of modified interaction potential, which contains additional soft repulsion.

\section{A model} \label{representation_GPF}

Consider a system of $N$ interacting particles located in the volume $V$. The potential of interaction is a function of distance $r$ between particles 
\begin{align}\label{1d1}
 &	U(r) = C_H \left\{ Ae^{-n_0(r-R_0)/\alpha} + e^{-\gamma(r-R_0)/\alpha} - 2e^{-(r-R_0)/\alpha}\right\}.
\end{align}
In the latter expression $R_0$ is the equilibrium distance between two particles (location of the minima of the potential), $\alpha$ is the effective radius of interaction, $\gamma, n_0$ are some parameters of the model, $C_H$ and $A$ are the normalizing constants. Location of the minima of the function $U(r)$ at $r = R_0$ is to be found from the following condition
\be\label{1d2}
\frac{\partial U(r)}{\partial r}\Bigg| = 0 \Rightarrow A = \frac{2-\gamma}{n_0}.
\ee
The depth of the potential well, which, according to \cite{ref8}, is equal to the energy of dissociation $ D $
\be\label{1d2a}
U(R_0) = - D .
\ee
The value of the constant $C_H$ is determined from the condition \eqref{1d2a}
\be\label{1d3}
C_H = D \frac{n_0}{n_0+\gamma - 2}.
\ee
Obviously the interaction potential \eqref{1d1} has two parameters $ n_0 $ and $ \gamma $, since  characteristic quantities $ R_0 $, $ D $, and $ \alpha $ of a specific physical system are known from numerical computations and their comparison with experiments. In particular, for sodium (Na) \cite{ref8} we have
\be\label{1d4}
R_0 = 5.3678, \quad 1/\alpha = 0.5504, \quad R_0/\alpha = 2.9544.
\ee
The constant of dissociation energy $ D $ for Na takes on a value \cite{ref9}
\[
D = 0.9241 \cdot 10^{-13} ergs.
\]
Note that in case of $n_0\gg 1$ the former term in \eqref{1d1} corresponds to the hard core (for any $r<R_0$ it goes to infinity and for $r>R_0$ it turns into zero). The case of $C_H\rightarrow D$ when $A\rightarrow 0$ yields the ordinary Morse potential \cite{ref8}.

To write a lattice form of the Morse potential \eqref{1d1} use the following notations
\begin{align}\label{1d5}
	& \tilde\Phi^{(r)}(r) = C_H e^{-\gamma(r-R_0)/\alpha}, \non
	& \Phi^{(a)}(r) = 2 C_H e^{-(r-R_0)/\alpha}, \non
	& \tilde \Psi(r) = C_H A e^{-n_0(r-R_0)/\alpha}.
\end{align}
Then define $l_{12}=|\vl_1-\vl_2|$ as the distance between two cells $l_1$ and $l_2$. According to \cite{ref1} the volume of the system is conditionally divided into $N_v$ cubic cells such that $V=c^3\cdot N_v$, where $c$ is the side and $v=c^3$ is the volume of each cell. The set of cell vectors is defined as
\begin{align}\label{1d6}
	&\Lambda = \Big\{  \vl = (l_x, l_y, l_z)|l_i = c\cdot n_i; \; n_i = 1,2,..., N_i; i = x,y,z; \; N_i= N_v^{1/3}\Big\}.
\end{align}
In thermodynamic limit $V\rightarrow \infty$, $N_v\rightarrow \infty$, and $v= V/N_v = const$
\begin{align}\label{1d7}
	& \tilde\Phi^{(r)}_{l_{12}} = C_H e^{-\gamma(l_{12}-c)/\alpha_R c}, \non
	& \Phi^{(a)}_{l_{12}} = 2 C_H e^{-(l_{12}-c)/\alpha_R c}, \non
	& \tilde \Psi_{l_{12}} = C_H A e^{-n_0(l_{12}-c)/\alpha_R c}.
\end{align}
Here $\alpha_R = \alpha/R_0$ is a dimensionless quantity. Consider $l_{12}=x\cdot c$, then the expressions \eqref{1d7} have the following form
\begin{align}\label{1d8}
	& \tilde\Phi^{(r)}(x) = C_H e^{-\gamma(x-1)/\alpha_R}, \non
	& \Phi^{(a)}(x) = 2 C_H e^{-(x-1)/\alpha_R}, \\
	& \tilde \Psi(x) = C_H A e^{-n_0(x-1)/\alpha_R}.
\end{align}
The latter expressions coincide with \eqref{1d5} if one takes into account that $r = x \cdot R_0$, $\alpha_R = \alpha / R_0$.
The Fourier transforms of the interaction potentials \eqref{1d7} are as follows
\begin{align}\label{1d9}
	& \Phi^{(r)}(k) = C_H 8\pi e^{\gamma/\alpha} \left( \frac{\alpha}{\gamma}\right)^{\!\! 3} \left( 1 + \left( \frac{\alpha}{\gamma}\right)^{\!\! 2} k^2 \right)^{\!\! -2}  \!\!\!\!,\non
	& \Phi^{(a)}(k) = C_H 16\pi e^{1/\alpha} \alpha^3 \left( 1 + \alpha^2 k^2 \right)^{-2} \!\!\!\!,\\
	& \Psi(k) = C_H A 8\pi e^{n_0/\alpha} \left( \frac{\alpha}{n_0}\right)^{\!\! 3} \left( 1 + \left( \frac{\alpha}{n_0}\right)^{\!\! 2} k^2 \right)^{\!\! -2} \!\!\!\!. \nonumber
\end{align}
Here and forth to simplify notations $\alpha$ denotes the quantity $\alpha_R$. Easy to see that
\begin{align} \label{1d10}
	& \Phi^{(a)}(0) = B \Phi^{(r)} (0), \quad B = 2 \gamma^3 e^{(1-\gamma)/\alpha},\non
	& \Psi(0) = A_\gamma \Phi^{(r)} (0), \quad A_\gamma = A e^{(n_0-\gamma)/\alpha} \left( \gamma / n_0\right)^3.
\end{align}
For the values \eqref{1d4} of the parameters $ R_0 $ and $ \alpha $, the quantity $ B $ expressed in \eqref{1d10} is small for the large values of $\gamma$ $(\gamma\gg 1)$, moreover $B\geq 1$, when $\gamma\leq \gamma_0$, where $\gamma_0=1.87047$. The behavior of the coefficient  $B(\gamma)$ is important in the subsequent calculations, in particular when determining the relation between the Fourier transform of repulsive  $\Phi^{(r)}(0)$ and the attractive $\Phi^{(a)}(0)$ interaction. In the range $\gamma<\gamma_0$ one has
\[
\Phi^{(a)} (0) > \Phi^{(r)}(0).
\]
This condition allows us to take advantage of the results of \cite{ref7}, where we obtained the following expression for the grand partition function of a cell fluid model
\begin{align}\label{1d11}
 &	\Xi = \int (d\rho)^{N_v} \exp \left[ \beta \tilde\mu \rho_0 + \frac{\beta}{2} \sum_{\vk\in B_c} W(k) \rhok \rhomk \right] \prod_{l=1}^{N_v} \left( \sum_{m=0}^{\infty} \frac{(\alpha^*)}{m!} e^{-pm^2} e^{m\tilde t_{\vl}} \right).
\end{align}
Here $\alpha^*=v e^{\beta_c\mu^*}$, $\tilde\mu=\mu-\mu^*(1+\tau)$. The effective interaction potential in case of the modified Morse potential is as follows
\begin{align}\label{1d12}
 & 	W(k)= \Phi^{(a)}(k) - \Phi^{(r)}(k) - \Psi(k) + \frac{\beta_c}{\beta} \chi_0 \Phi^{(r)}(0) + \frac{\beta_c}{\beta} \Psi(0).
\end{align}
Here $\chi_0$, $\mu^*$ are some constants, and $\tau$ is the relative temperature
\be\label{1d13}
\tau = (T - T_c) / T_c,
\ee
here $T_c$ is the critical temperature which is to be defined later. Easy to see, that
\be\label{1d14}
W(0) = \Phi^{(r)}(0) \left[ B - 1 + \chi_0 + \tau (\chi_0 + \epsilon) \right].
\ee
The condition $W(0) > 0$ is met for all $\gamma < \gamma_0$.
The parameter $p$ from \eqref{1d11} has the form
\be\label{1d15}
p = \frac{\beta_c}{2} \Phi^{(r)}(0) [\chi_0 + A_\gamma].
\ee

Expressing the last factor in \eqref{1d11} as the cumulant series gives \cite{ref7}
\begin{align}\label{1d16}
	\Xi & = g_v \int (d t)^{N_v} \exp \left[ - \frac{1}{2} \sum_{\vk\in B_c} t_{\vk} t_{-\vk}/\beta W(k) \right] \prod_{l=1}^{N_v} \exp \left( \sum_{m=0}^{\infty} \frac{g_n}{n!} \tilde t_{\vl}^n \right).
\end{align}
In the latter expression, the Stratonovich-Hubbard transform has been already used for the factor in \eqref{1d11} containing the effective interaction potential, since $W(k)>0$, as follows with \eqref{1d12}. The quantity $g_v$ in \eqref{1d16} is denoted by the formula
\[
g_v = \prod_{\vk\in B_c} (2\pi \beta W(k))^{-1/2},
\]
and the variable $\tilde t_{\vk}$ is
\be\label{1d18}
\tilde t_{\vk} = t_{\vk} + \beta_c \tilde\mu \sqrt{N_v} \delta_{\vk},
\ee
moreover,
\[
\tilde t_{\vl} = \frac{1}{\sqrt{N_v}} \sum_{\vk\in B_c} \tilde t_{\vk} e^{-i\vk\vl}.
\]
The coefficients $g_n$ from \cite{ref7,ref3} are as follows
\begin{align}\label{1d19}
	& g_0 = \ln T_0, \quad g_1 = T_1/T_0, \quad g_2 = T_2/T_0 - g_1^2,  \non
	& g_3 = T_3/T_0 - g_1^3 - 3g_1 g_2,\\
	& g_4 = T_4/T_0 - g_1^4 - 6 g_1^2 g_2 - 4 g_1 g_3 - 3 g_2^2,\nonumber
\end{align}
where $T_n(p,\alpha^*)$ are the following special functions
\be\label{1d20}
T_n(p,\alpha^*) = \sum_{m=0}^{\infty} \frac{(\alpha^*)^m}{m!} m^n e^{pm^2}.
\ee
Since $p$ fails to be the function of temperature, the coefficients of the Jacobian of transition to the collective variables $g_n$ don't depend on $\tau$.

In an approximation of the $\rho^4$ model one has \cite{ref11}
	\be \label{1d21}
	\Xi = g_V  e^{N_vE_\mu} \int (d\rho)^{N_v} \exp \left[ M N_v^{1/2} \rho_0 + \frac{1}{2} \sum_{\vk\in\cB_c} \tilde D(k) \rhok\rhomk 
	+  \frac{g_4}{24} \frac{1}{N_v} \sum_{\substack{\vk_1,...,\vk_4 \\ \vk_{i}\in \cB_c}}  \rho_{\vk_1}...\rho_{\vk_4} \delta_{\vk_1+...+\vk_4}\right].
	\ee 
The following notations are used in the latter formula 
\begin{align}\label{1d22}
	& E_\mu = g_0 - \frac{\beta\tilde\mu^2}{2W(0)} + n_c \! \left( \! g_1 \! + \frac{\tilde\mu}{W(0)} \! \right) \! + \frac{n_c^2}{2} \tilde D(0) + \frac{g_3^4}{8g_4^3},\non
	& M = \tilde\mu / W(0) + g_1 + n_c \tilde D(0) - \frac{1}{6} \frac{g_3^3}{g_4^2},\non
	& \tilde D(k) = \tilde g_2 - 1 /\beta W(k), \\
	& \tilde g_2 = g_2 - \frac{1}{2} \frac{g_3^2}{g_4}, \quad n_c = - g_3 / g_4. \nonumber
\end{align}
The expression \eqref{1d21} is approximated only in the sense that it takes into account contribution from cumulants $ g_n $ of $ n \leq 4 $. Higher $ g_n $ (with $ n \geq 5 $) are considered to be zero. As shown in \cite{ref10}, this approximation qualitatively describes the behavior of a three-dimensional Ising model near a critical point. Since simple fluids belong to the universality class of the Ising model, hopefully, the approximation used in \eqref{1d21}  will allow us to describe the first-order phase transition in simple fluids, including the critical region.

\section{Temperature dependence and the range of model parameters}

To compute the values of $ g_n $ that are required to obtain the explicit form of \eqref{1d21}, it is essential to have the parameters $ p $ and $ \alpha^* $. Consider the easiest case when
\be\label{2d1}
p = 1.0, \quad \alpha^* = 5.0
\ee
The value of $ p $ is unambiguously determined by the parameters $ \chi_0 $ and $ A_\gamma $, which are part of \eqref{1d15}. The value of $ A_\gamma $ is a function of the parameter $ n_0 $ defined in \eqref{1d10}, so the case of $ p = 1 $ correspond to the well-defined $ \chi_0 $ and $ n_0 $. Let us find their magnitudes. 
Conforming to \eqref{1d10} if
\begin{equation}\label{2d1a}
	\gamma = 1.650,
\end{equation}
then
\[
n_0 = 1.521, \quad A_{\gamma} = 0.201; \quad \chi_0 = 0.070.
\]

Examine \eqref{1d21} applying the zero-mode approximation, just as it had been done in \cite{ref7}. The result is a grand partition function in the following form
\be\label{2d2}
\Xi = g_v N_v^{1/2} \exp [N_v E(\bar\rho_0)],
\ee
where
\be\label{2d3}
E(\bar \rho_0) = M \bar\rho_0 + \frac{1}{2} \tilde D(0) \bar\rho_0^2 - \frac{a_4}{24} \bar\rho_0^4,
\ee
moreover the coefficient $a_4 = -g_4>0$. The quantity $\bar\rho_0$ is a solution of the equation $\partial E(\rho_0) / \partial \rho_0 = 0$, which is
\be\label{2d4}
\bar M + \tilde D(0) \bar\rho_0 - \frac{a_4}{6} \bar\rho_0^3 = 0,
\ee
here $\bar M$ is some value of the chemical potential which corresponds to the extrema of function $E(\rho_0)$. This extremum is the maximum (which is a condition of the Laplace method used to calculate \eqref{2d2}) if the inequality
\be\label{2d5}
E_2 (\bar\rho_0) \leq \frac{\partial^2 E(\bar\rho_0)}{\partial \rho_0^2}
\ee
is fulfilled. The equality condition in \eqref{2d5} corresponds to the quantity
\be\label{2d6}
\rho_{0r}^{2} = 2 \frac{\tilde D(0)}{a_4}.
\ee
Later we will see that $ \rho_{0r} $ describes a spinodal curve that is specified only in range $ T <T_c $.

The critical temperature $T_c$ is provided by solving the equation
\be\label{2d10}
\tilde D(0, T_c) = 0.
\ee
For the parameter values given by \eqref{2d1} and \eqref{2d1a} one has
\begin{align}\label{2d11}
	& k_B T_c = \tilde g_2 W(0, T_c) = \tilde g_2 (B - 1 + \chi_0) \Phi^{(r)}(0), \\ 
	& k_B T_c \approx 4.995. \nonumber
\end{align}

Find the explicit form of the dependence of $ \tilde D (0) $ on temperature, based on \eqref{1d22}. In view of \eqref{1d14}, one has
\be\label{2d7}
\frac{1}{\beta W(0)} = \tilde g_2 \cdot \gamma_\tau, \quad \gamma_\tau = \frac{1+\tau}{1+\omega_0\tau},
\ee
where $\tilde g_2$ is an independent of temperature constant defined by the microscopic parameters \eqref{1d4} and \eqref{2d1}, 
\be\label{2d8}
\omega_0 = \frac{\chi_0 + A_\gamma}{B - 1 + \chi_0}.
\ee
Formula \eqref{1d22} provides the following expression
\be\label{2d9}
\tilde D(0) = - \tau \frac{1-\omega_0}{1+\omega_0\tau} \tilde{g}_2.
\ee
It comes from the computation of the coefficients $g_n$ given by \eqref{1d19} that under conditions \eqref{2d1} the value $\tilde g_2<1/2$, therefore $\omega_0 = 2 \tilde g_2<1$. Consequently, the quantity $\tilde D(0)\leq 0$ for all $\tau\geq 0$, and $\tilde D(0)\geq 0$ in case $\tau\leq 0$. This leads us to the conclusion that in the temperature range $ \tau> 0 $ the equation \eqref{2d6} has no real solutions for $ \rho_{0r} $, and therefore there are no restrictions on the quantity $ \bar \rho_0 $ as a solution of the equation \eqref{2d4}. However, in case $ \tau <0 $ there is a real quantity
\be\label{2d14}
\rho_{0r} =  \left( 2 \frac{\tilde D(0)}{a_4} \right)^{1/2}
\ee
as well as the following restriction on the value of $\bar \rho_0$
\be\label{2d15}
|\bar \rho_0|\geq \rho_{0r}.
\ee
The equality sign in \eqref{2d15} matches a spinodal curve that limits the stability region of the system.

The reduced form of the equation \eqref{2d4} is given by
\be\label{2d16}
\bar \rho_0^3 + p_t \bar \rho_0 + q = 0,
\ee
where
\be\label{2d17}
p_t = - \frac{6 \tilde D(0)}{a_4}; \quad q = - \frac{6\bar{M}}{a_4}.
\ee
The solution of \eqref{2d16} in the range $T>T_c$ provided by
\be\label{2d18}
\bar \rho_0 = \lp \frac{3\bar{M}}{a_4} + \sqrt{Q_t}\rp^{1/3} - \lp - \frac{3\bar{M}}{a_4} + \sqrt{Q_t} \rp^{1/3} \!\!\!\! ,
\ee
where
\be\label{2d19}
Q_t = \lp - \frac{2\tilde D(0)}{a_4}\rp^3 + \lp \frac{3\bar{M}}{a_4}\rp^2.
\ee
For all $T>T_c$ the quantity $p_t>0$. That suggests the existence of a unique real solution of \eqref{2d16}. Using both a well-known relation
\be\label{2d20}
PV = k_B T \ln \Xi
\ee
and the expression \eqref{2d2} gives the following explicit form of the equation of state in  case $T>T_c$ 
\be\label{2d21}
\frac{PV}{k_B T} = \frac{1}{N_v} \ln g_v + E_\mu(T) + \bar{M} \bar \rho_0 + \frac{1}{2} \tilde D(0) \bar \rho_0^2 - \frac{a_4}{24} \bar \rho_0^4.
\ee
The quantity $E_\mu(T)$ is provided by

\begin{align}\label{2d22}
	& E_\mu = g_0 - \frac{\beta W(0)}{2} \!\! \lp \! \frac{\tilde\mu}{W(0)} \! \rp^{\! 2} \!\!\! + n_c \bar{M} \! - \frac{n_c^2}{2}  \tilde D(0) - \frac{1}{24} \frac{g_3^4}{g_4^3}, \non
	& \frac{\tilde\mu}{W(0)} = \bar{M} - g_1 - n_c \tilde D(0) + \frac{1}{6} g_3^3 / g_4^2,
\end{align}
Note that for all $ \tau> 0 $ there are no restrictions on the value of $ \bar \rho_0 $. The expression \eqref{2d21} for pressure is a monotonically increasing function of chemical potential and temperature.

\section{Region of temperatures below the critical one}

For all $\tau<0$ the equation \eqref{2d16} has three real solutions
\begin{align} \label{4d1}
	& \rho_{01} = 2 \rho_{or} \cos (\alpha_t / 3), \non
	& \rho_{02} = - 2 \rho_{or} \cos \lp \alpha_t / 3 + \frac{\pi}{3}\rp, \\
	& \rho_{03} = - 2 \rho_{or} \cos \lp \alpha_t / 3 - \frac{\pi}{3}\rp, \nonumber
\end{align}
where $\rho_{0r}$ is given by \eqref{2d14}. The quantity $\alpha_t$ is provided by
\be\label{4d3}
\cos \alpha_t = \bar M / M_q,
\ee
moreover
\be\label{4d4}
M_q = - \frac{g_4}{3} \rho_{0r}^3 = \frac{a_4}{3} \lp \frac{2 \tilde D(0)}{a_4}\rp^{3/2}.
\ee
Note that the solutions \eqref{4d1} are valid in case of $Q_t<0$, where $Q_t$ is defined in \eqref{2d19}, that is for all $|\bar M|<M_q$, where $M_q$ is the solution of equation $Q_t=0$.
\begin{figure}[h!]
	\centering \includegraphics[width=0.5\textwidth]{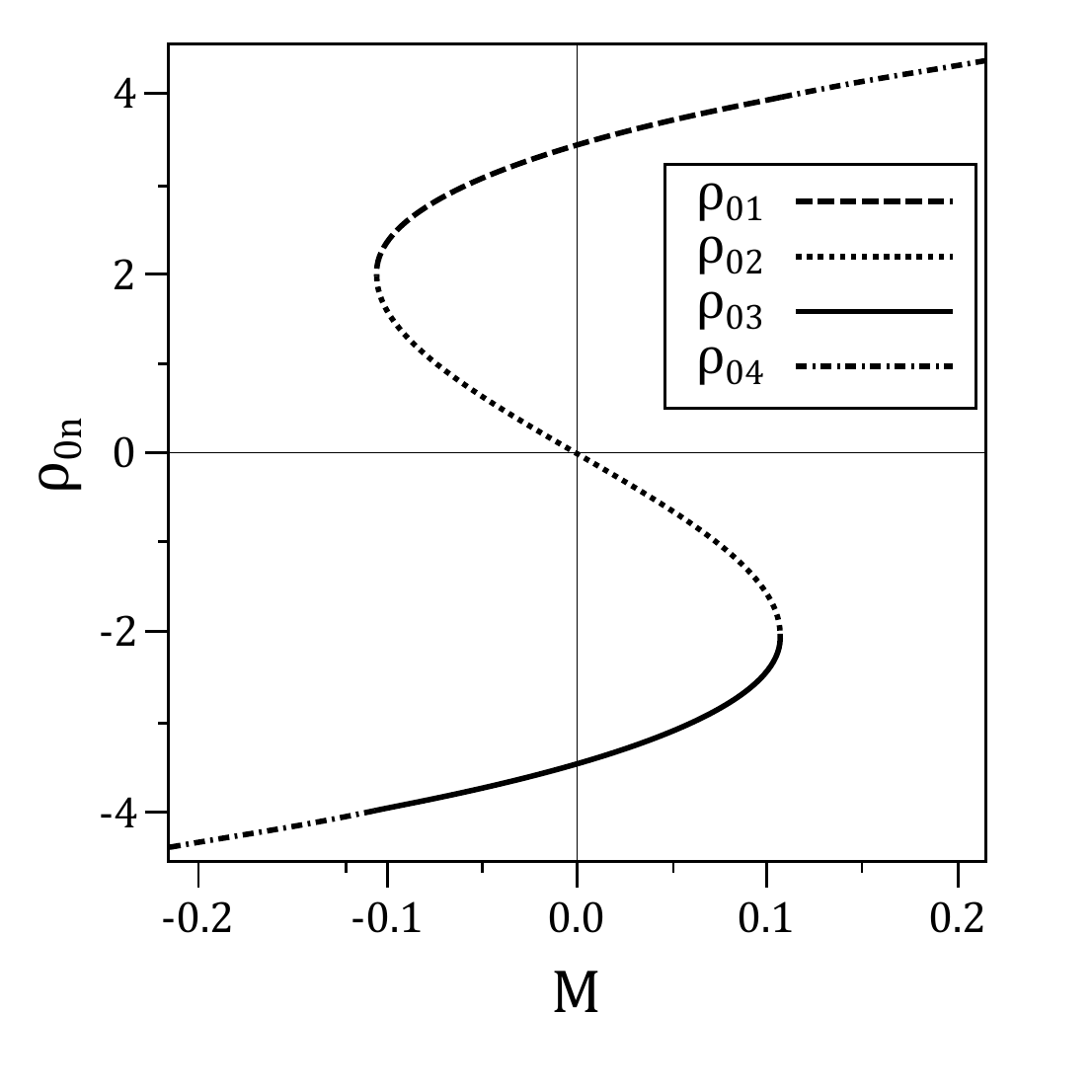}
	\vskip-3mm\caption{Plot of solutions $\bar \rho_0$ (see \eqref{4d1}) at $\tau = - 0.5$. In the range $Q_t<0$ there exist three real roots, at $Q_t>0$ \--- a unique real solution.
	}\label{fig1}
\end{figure}
A unique real solution $\bar \rho_0$ of the equation  \eqref{2d16} exists in the range $|M|\geq M_q$. It is calculated similarly to \eqref{2d18} because for these values of $ M $ the quantity $ Q_t \geq 0 $:
\be\label{4d5}
\rho_{04} = \left(\frac{3}{a_4}\right)^{\frac{1}{3}} (A_1^{\frac{1}{3}} + A_2^{\frac{1}{3}}),
\ee
where
\begin{align}\label{4d6}
	& A_1 = \bar M + \sqrt{\bar M^2 - M_q^2}, \non
	& A_2 = \bar M - \sqrt{\bar M^2 - M_q^2}.
\end{align}
The solution $\rho_{04}$ defined in \eqref{4d5} is an extension of one of the solutions \eqref{4d1}. Moreover for all $\bar M>0$ $\rho_{04}$ is the continuation of the solution $\rho_{01}$, and in range $\bar M<0$ it coincides with $\rho_{03}$ at $\bar M=-M_q$. The corresponding curves are shown in Figure~\ref{fig1}.

\begin{figure}[h!]
	\centering\includegraphics[width=0.6\textwidth]{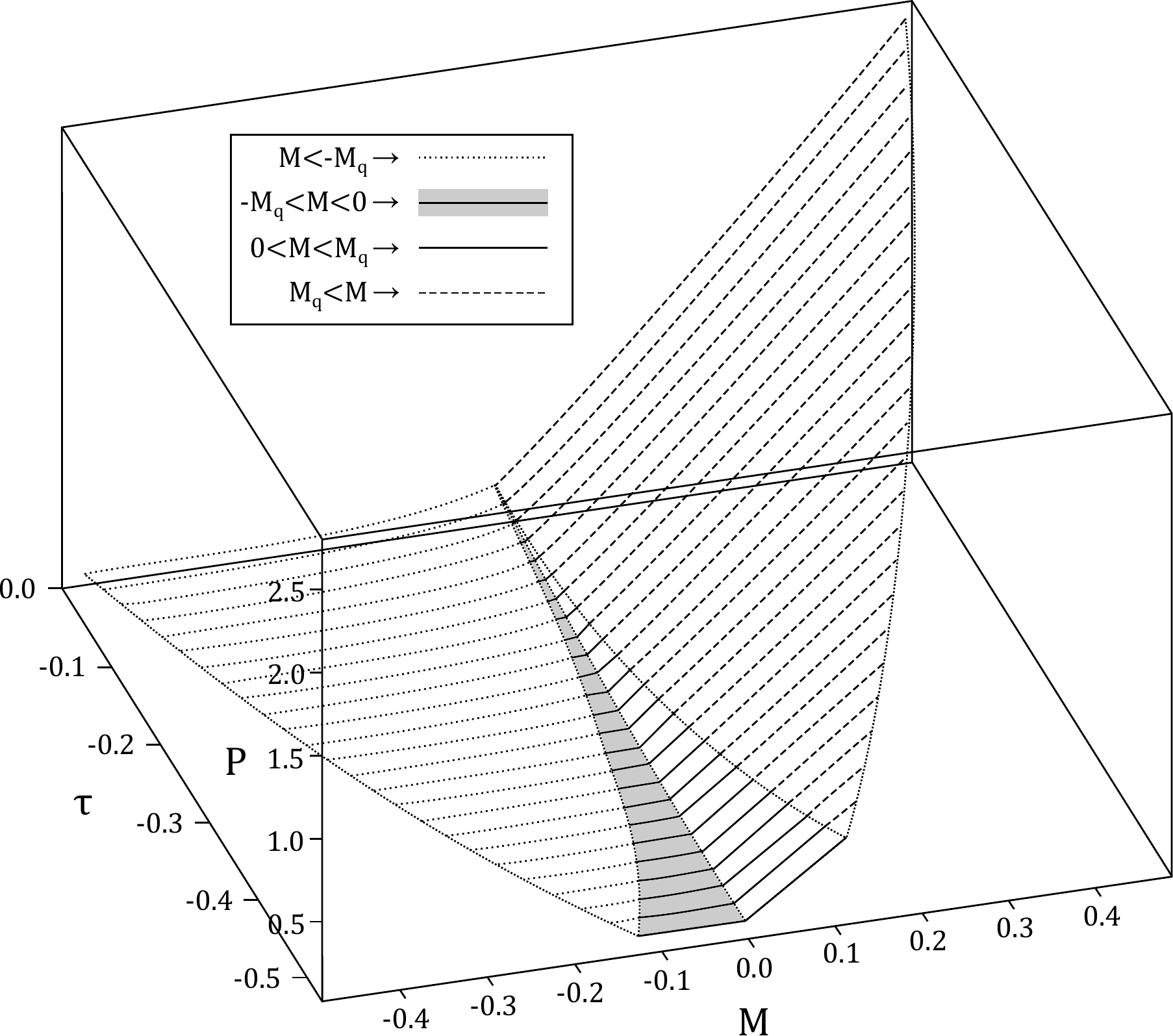}
	\vskip-3mm\caption{3D plot of pressure as a function of temperature and chemical potential ($P=P(T,M)$) in range $T<T_c$.}\label{fig2}
\end{figure}

Note that
\begin{align}\label{4d7}
	\lim_{\bar M\rightarrow M_q} \rho_{01} & = \lim_{\bar M\rightarrow M_q} \rho_{04}^{(p)}  = 2 \rho_{0r} \lim_{\bar M\rightarrow - M_q} \rho_{03} = \non  
	& = \lim_{\bar M\rightarrow - M_q} \rho_{04}^{(m)} = - 2 \rho_{or},
\end{align}
where $\rho_{04}^{(p)}$ and $\rho_{04}^{(m)}$ correspond to \eqref{4d5} in either the range $\bar M>0$ or $\bar M<0$ respectively.
An equation of state at $T<T_c$ in terms of chemical potential and temperature has the following form
	\begin{align}\label{4d11}
		{\frac{PV}{k_B T}}  &= \frac{1}{N_v} \ln g_V + E_\mu(\mu, T) + E(\rho_{01}) \Theta(M) \Theta(M-M_q) + E(\rho_{03}) \Theta(-M) \Theta(M_q-M) +  \non
		& + E(\rho_{04}^{(m)}) \Theta(-M-M_q) + E(\rho_{04}^{p}) \Theta(M-M_q),
	\end{align}
where the function $E(\rho_0)$ is specified in \eqref{2d3}, and the arguments of this function are expressed by \eqref{4d1} and \eqref{4d5}. The quantity $E_\mu(\mu,T)$ is given by \eqref{2d22}. The 3D plot of pressure below the $T_c$ is represented in Figure~\ref{fig2}.

\section{Equation of relation between the chemical potential and the density}

The above calculated expressions for the equation of state \eqref{2d21} at $ T> T_c $ as well as \eqref{4d11} at $ T <T_c $ contain a pressure dependence on temperature and chemical potential. To obtain the pressure as a function of temperature and density, we use the known relation for the average number of particles
\be\label{3d1}
<N> = \frac{\partial}{\partial\beta\mu} \ln \Xi .
\ee
Taking into account the expression \eqref{2d2} of $\Xi$, one has
\be\label{3d2}
\bar n = \frac{<N>}{N_v} = \frac{\partial E_\mu}{\partial\beta\mu} + \frac{\partial E(\bar \rho_0)}{\partial \beta \bar M}.
\ee
The following equation which links the density of particles $\bar n$ and the chemical potential $\bar{M}$ is obtained using the formulas \eqref{2d3} and \eqref{2d22}
\be\label{3d3}
\bar n = n_g - \bar{M} + \tilde g_2 \gamma_\tau \bar \rho_0,
\ee
where
\be\label{3d4}
n_g = g_1 + n_c \tilde g_2 - \frac{1}{6} g_3^3 / g_4^2.
\ee
Rewrite \eqref{3d3} in the form
\be\label{3d5}
\bar \rho_0 = \lp \bar M + \bar n - n_g \rp / \tilde g_2 \gamma_\tau.
\ee
The latter expression is actually being an equation for $\rho_0$, since $\bar M$ contained in \eqref{3d3} is the function of the quantity $\bar \rho_0$. This follows from the equality \eqref{2d4}, which specifies the extreme value of $\bar \rho_0 (\tau, \bar M)$. Substituting $ \bar {M} $ from \eqref{2d4} to \eqref{3d5}, we obtain the equation for $\rho_{0n} = \rho_{0n} (\tau, \bar n)$.
\be\label{3d8}
\rho_{0n}^3 + p_q \rho_{0n} + q_q = 0,
\ee
where the coefficients $p_q$ and $q_q$ are given by
\be\label{3d9}
p_q = - \frac{6}{a_4} (\tilde g_2 \gamma_\tau + \tilde D(0)); \quad q_q = \frac{6}{a_4} (\bar n - n_g).
\ee
The equation \eqref{3d8} allows us to find $ \bar \rho_0 $ as a function of density $ \bar n $ and temperature. It transforms the solution of the equation \eqref{2d16} from the dependence on chemical potential to density dependence.

\begin{figure}[h!]
	\centering\includegraphics[width=0.5\textwidth]{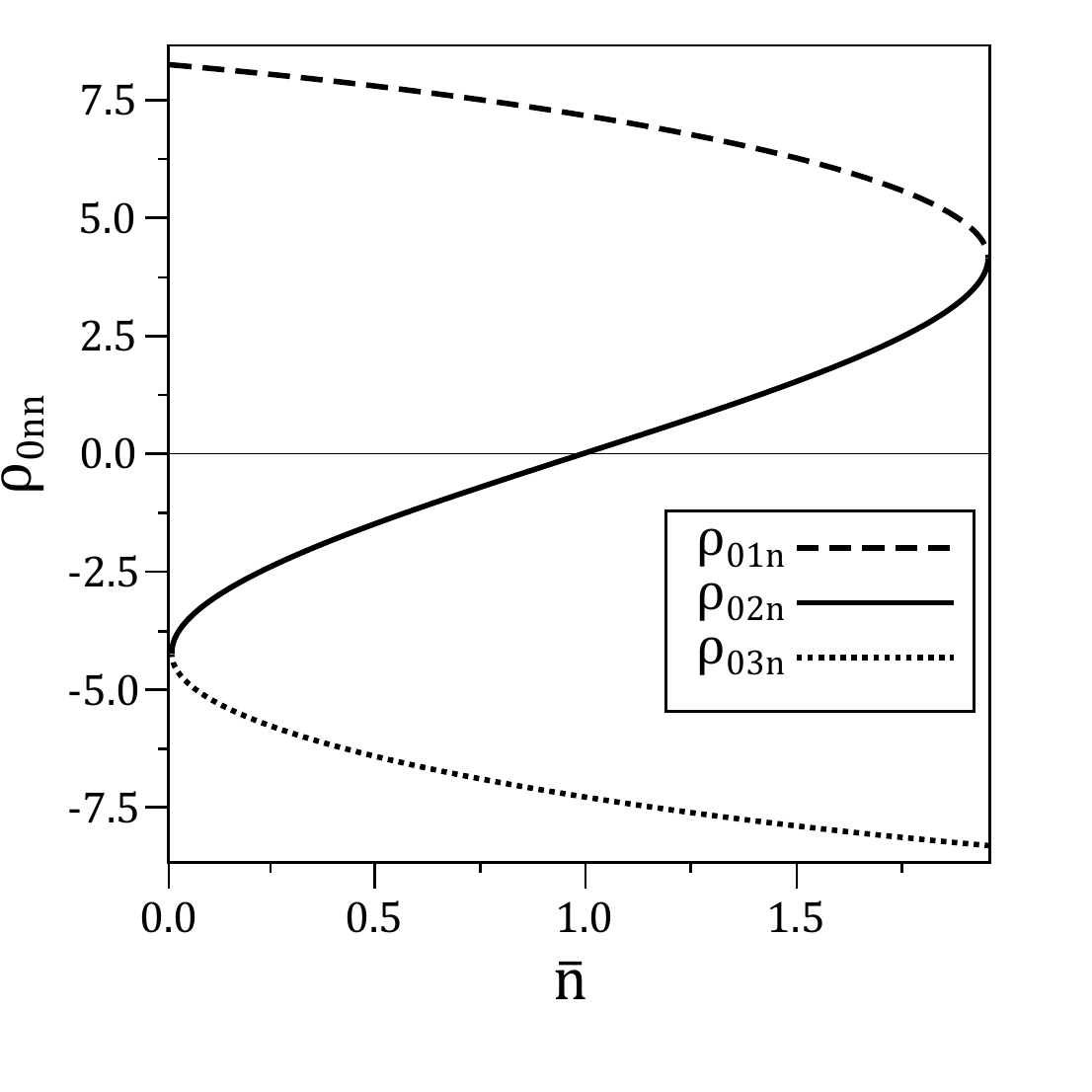}
	\vskip-3mm\caption{Plot of solutions $\rho_{0nn}$ given by \eqref{3d12} as functions of density.
	}\label{fig3}
\end{figure}

Easy to see that in keeping with \eqref{2d7} and \eqref{2d9} the coefficient $p_q$ is a constant.
\be\label{3d10}
p_q = - \frac{6\tilde g_2}{a_4}.
\ee
The latter is negative and independent on temperature. Therefore, for all $Q_q<0$ given by 
\be\label{3d11}
Q_q = (p_q/3)^3 + (q_q/2)^2.
\ee
the equation \eqref{3d8} possesses three real solutions
\begin{align} \label{3d12}
	& \rho_{01n} = 2 \lp \frac{2\tilde g_2}{a_4} \rp^{1/2} \cos \lp \alpha_n / 3 \rp, \non
	& \rho_{02n} = - 2 \lp \frac{2\tilde g_2}{a_4} \rp^{1/2} \cos \lp \alpha_n / 3 + \pi / 3\rp, \\
	& \rho_{03n} = - 2 \lp \frac{2\tilde g_2}{a_4} \rp^{1/2} \cos \lp \alpha_n / 3 - \pi / 3\rp. \nonumber
\end{align}
Here either
\[
\cos \alpha_n =  \frac{n_g - \bar n}{n_\varphi}; \quad n_\varphi = \frac{2}{3} \lp \frac{2\tilde g_2^3}{a_4}\rp^{1/2},
\]
or
\be\label{3d13}
\alpha_n = \arccos \left( \frac{n_g - \bar n}{n_\varphi} \right) .
\ee
The solutions \eqref{3d12} are shown on figure~\ref{fig3}. Extension of \eqref{3d8} to the range $Q_q>0$ leads to continuation of non-physical branches of the solutions $\rho_{03n}$ and $\rho_{01n}$ (see Figure~\ref{fig3}).

Note that both \eqref{2d16} and \eqref{3d8} are equations of the same type, but describe different dependencies. The former, namely \eqref{2d16}, allows us to find solutions $ \bar \rho_0 (\tau, M) $ of equation for finding the extrema of $ E (\bar \rho_0) $ provided by \eqref{2d3} according to application of the Laplace method \cite{ref16} for calculating the expression of the grand partition function $ \Xi (\tau, M) $. This is the way to obtain the equation of state \eqref{2d20} in terms of temperature and chemical potential. The latter equation --- \eqref{3d8} --- makes it possible to find both the quantity $ \bar \rho_0 $ and, consequently, the pressure as functions of density and temperature.

The difference between the equations \eqref{2d16} and \eqref{3d8} is also evident in the aspect of mathematics. Thus, the solutions $ \rho_{01} $ and $ \rho_{03} $ of the equation \eqref{2d16} corresponding to $ \max E (\bar \rho_0) $ are useful at $ T <T_c $. The quantity $ \rho_{02} $ refers to $ \min E (\bar \rho_0) $. Among the three solutions \eqref{3d8} only $ \rho_{02n} $ is physical, instead . This solution characterizes increasing density with the growth of chemical potential. The other two solutions --- $ \rho_ {01n} $ and $ \rho_ {03n} $ --- show a decrease in density (with growing chemical potential). Such a behavior does not fit the reality. The calculations should be carried out in the sequence suggested in the present paper. Firstly, find the solutions $ \bar \rho_0 (\tau, M) $ which give an explicit expression of the grand partition function $ \Xi (\tau, M) $ as a function of temperature and chemical potential. Only then search for $ \bar \rho_0 (\tau, \bar n) $. This procedure gives the way to obtain both $ \Xi (\tau, \bar n) $ and the equation of state $ P = P (\tau, \bar n) $ in terms of temperature and density.

\begin{figure}[h!]
	\centering\includegraphics[width=0.5\textwidth]{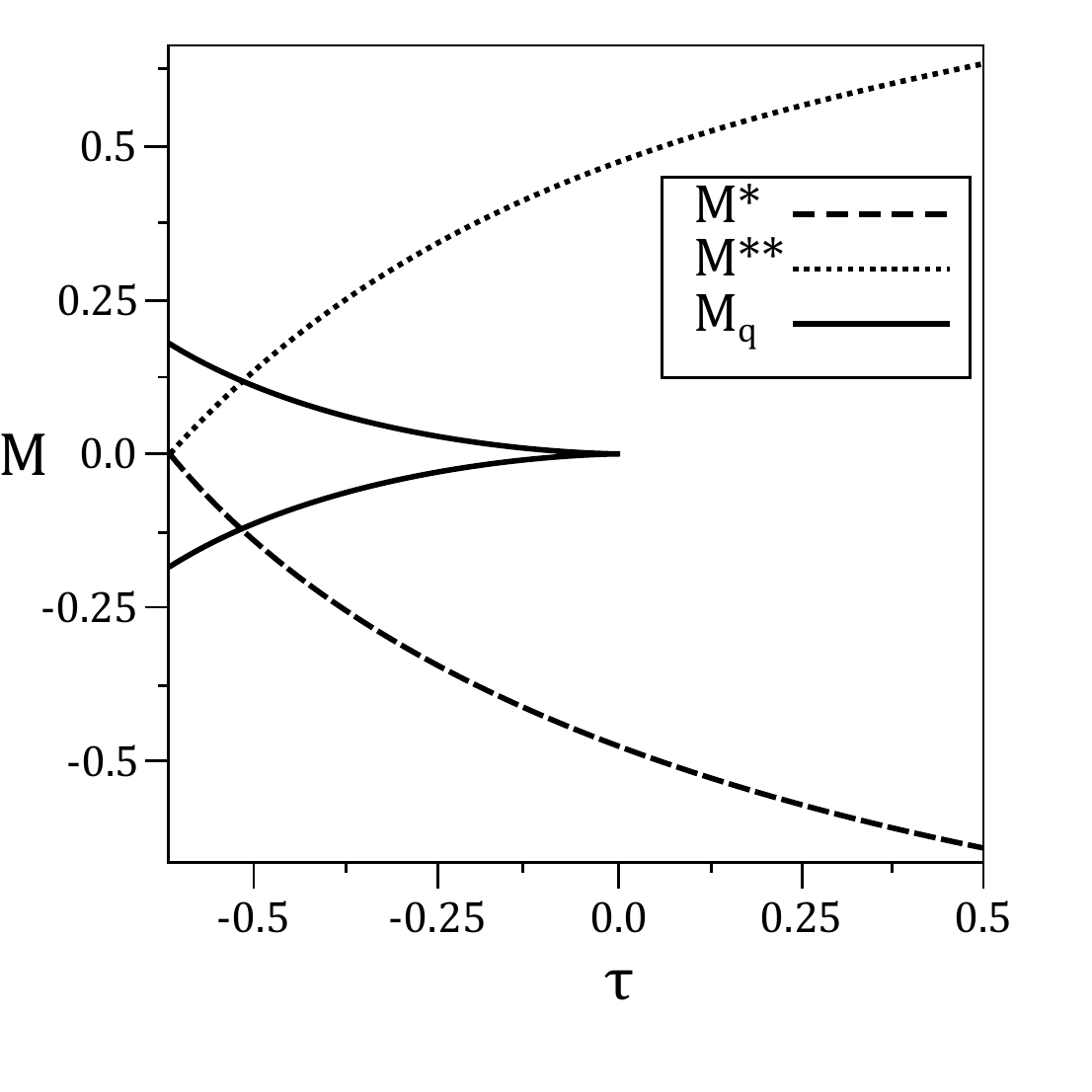}
	\vskip-3mm\caption{Boundary values of the chemical potential $M$ ($M_q$, $M^{**}$ --- the upper limit, $M^*$ --- the lower limit).}\label{fig4}
\end{figure}

Using a zero-mode approximation \eqref{2d2} of the $\rho^4$-model \eqref{1d21} leads to restriction on the range of the quantity $\rho_{0n} (\tau, \bar n)$ defined in \eqref{3d12} (see Figure~\ref{fig3})
\be\label{3d14}
- \bar \rho < \rho_{0n} < \bar \rho,
\ee
where
\be\label{3d15}
\bar \rho = \lp 2 \tilde g_2 / a_4 \rp^{1/2}.
\ee
Consider this situation in more details. For this purpose rewrite the equation \eqref{3d3} as follows
\be\label{3d16}
\bar M = \tilde g_2 \gamma_\tau \rho_{0n} - ( \bar n- n_g).
\ee
The quantity $\bar M$ (which is a variable in the framework of the grand canonical ensemble in terms of ($\tau, M$)) becomes a function of density $\bar n$ and temperature $\tau$. To transform the results from those represented in terms of ($\tau, M$) into $(\tau, \bar n)$ use the formula \eqref{3d16}, where the solution $\rho_{02n}$ from \eqref{3d12} is applied in $\rho_{0n}$. To simplify notations denote $\rho_{02n} = \rho_{n}$. Since the right-hand side of \eqref{3d16} is a finite quantity, the quantity $ \bar M $ is also limited by two boundary curves. The former corresponds to $ \rho_n = - \bar \rho $. Then \eqref{3d16} gives
\be\label{3d17}
M^* = \lim_{\rho_n\rightarrow - \bar \rho} \bar M = - \tilde g_2 \gamma_\tau \bar \rho - (n_{min} - n_g).
\ee
The following relation is used in the latter formula
\be\label{3d18}
\lim_{\rho_n\rightarrow - \bar \rho} \bar n = n_{min}.
\ee
The other limit provides an upper boundary curve, where $\rho_n \rightarrow \bar \rho$ and the density $\bar n \rightarrow n_{max}$
\be\label{3d19}
M^{**} = \lim_{\rho_n\rightarrow \bar \rho} \bar M = \tilde g_2 \gamma_\tau \bar \rho - ( n_{max}- n_g).
\ee
Here
\be\label{3d20}
\lim_{\rho_n\rightarrow  \bar \rho} \bar n = n_{max}.
\ee
Figure \ref{fig4} illustrates a temperature dependence of both $M^*$ and $M^{**}$.
Evidently, there is no gas-liquid phase transition below the value $n_{min}$ which specifies the lowest possible density for some temperature $ T^* $. Thus, in the approximation of the $\rho^4$-distribution, the cell fluid model has a certain range of admissible density values
\begin{equation}\label{3d18a}
	n_{min} \leq \bar n \leq n_{max}.
\end{equation}

Easy to see that there is some temperature
\begin{equation}\label{3d19a}
	\tau^* = -0.618,
\end{equation}
at which the following equality is fulfilled
\begin{equation}\label{3d20a}
	M^* = M^{**} = 0.
\end{equation}
This temperature corresponds to some fixed density values
\begin{align} \label{3d21}
	& n_{max} = n_g + n_\varphi,\non
	& n_{min} = n_g - n_\varphi,
\end{align}
where
\be\label{3d22}
n_\varphi = \bar \rho \tilde g_2 \gamma_\tau(\tau^*),
\ee
Clearly, there is a condition $n_{min} \geq 0$ so the following inequality should be held
\be\label{3d23}
n_g \geq n_\varphi.
\ee
This condition is met for the parameter set \eqref{2d1}. Use the parameters \eqref{2d1a} to describe sodium and find
\begin{align}\label{3d24}
	& n_g = 0.977; \qquad n_{max} = 1.946;\non
	& n_\varphi = 0.968; \qquad n_{min} = 0.009.
\end{align}
The quantity $n_{min}$ increases with decreasing parameter $\alpha^*$. For instance, $p = 1$ and $\alpha^* = 4$ give
\begin{align}\label{3d25}
	& n_g = 0.884; \qquad n_{max} = 1.652;\non
	& n_\varphi = 0.767; \qquad n_{min} = 0.117.
\end{align}
In case $p=1$, $\alpha^* = 3$:
\begin{align}\label{3d26}
	& n_g = 0.832; \qquad n_{max} = 1.505;\non
	& n_\varphi = 0.673; \qquad n_{min} = 0.158.
\end{align}
See that each set of parameters $ p $ and $ \alpha^* $ corresponds to particular values of characteristic densities.

\begin{figure}[h!]
  \begin{minipage}[t]{0.48\textwidth}
	\centering \includegraphics[width=\textwidth]{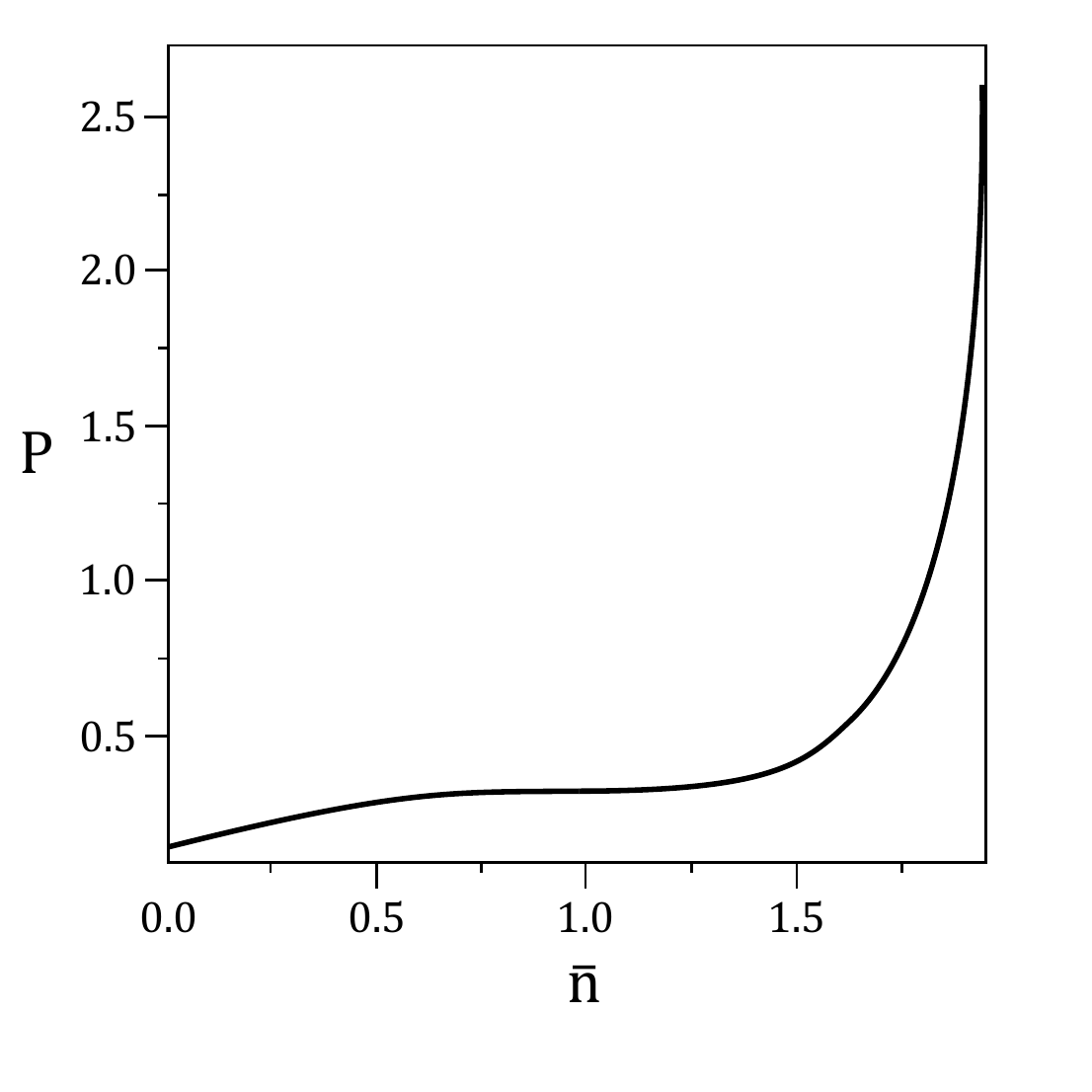}
	\vskip-3mm\caption{Plot of the pressure as a function of density at $T=T_c$.}\label{fig5}	
  \end{minipage}
  \begin{minipage}[t]{0.48\textwidth}
	\centering \includegraphics[width=\textwidth]{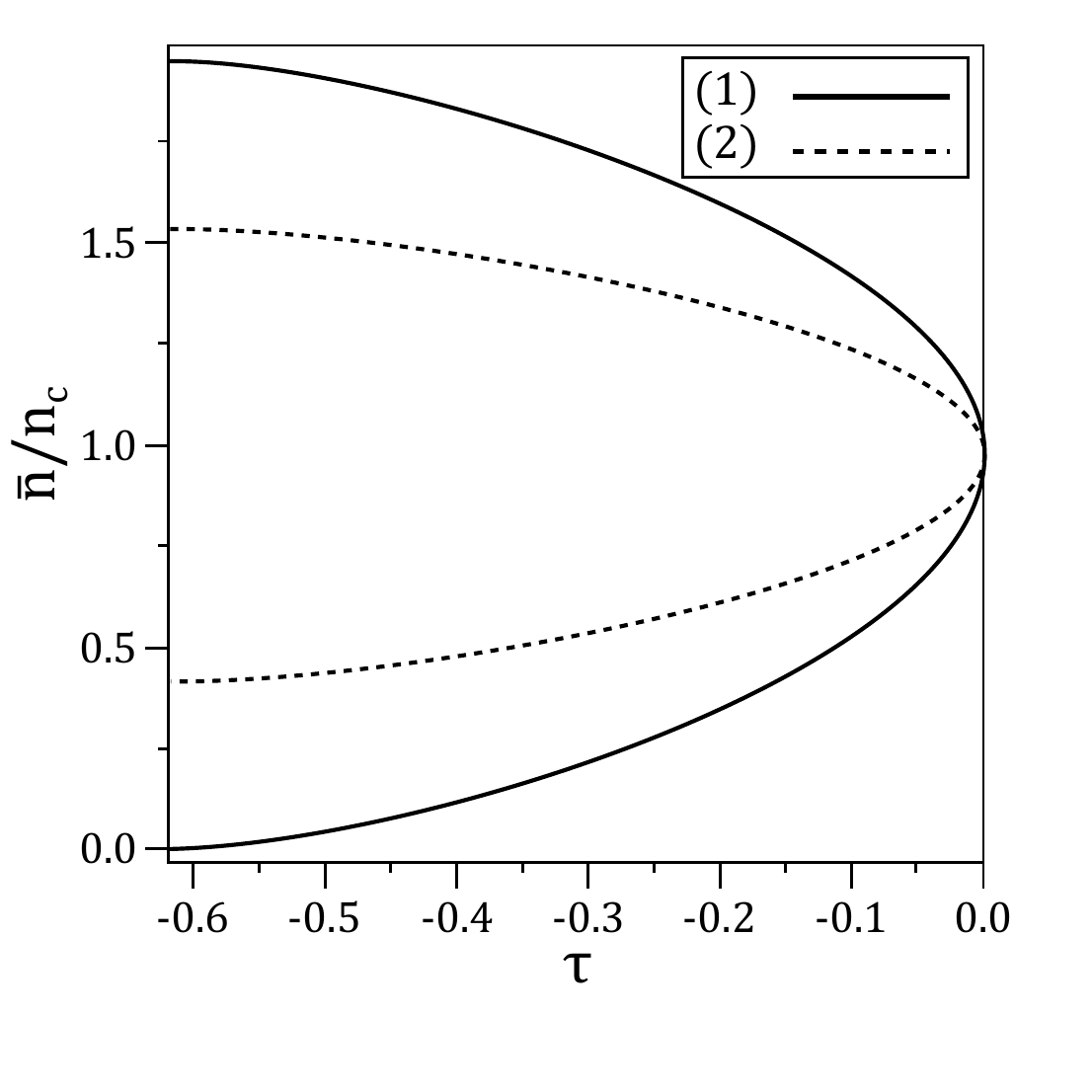}\\
	\vskip-3mm\caption{The coexistence curve (1) and the spinodal (2) of the model investigated in the present research.}\label{fig6}
  \end{minipage}
  \end{figure}

At $T>T_c$ the equation of state of a cell fluid model in terms of density has the following form
\be\label{3d29}
\frac{PV}{kT} = \frac{1}{N_v} \ln g_v + E_\mu^{(2)}(n,T) + \bar M \rho_{n} + \frac{\tilde D(0)}{2} \rho_{n}^2 - \frac{a_4}{24} \rho_{n}^4,
\ee
where
\be\label{3d30}
\bar M = \rho_{n} \tilde g_2 \gamma_\tau - (\bar n - n_g)
\ee
is a function of density and temperature. Recall the formula \eqref{2d22} to find
\be\label{3d31}
\frac{\tilde\mu}{W(0)} = \bar M - n_g + n_c \tilde g_2 \gamma_\tau,
\ee
where $n_g$ is defined in \eqref{3d4}. Therefore,
\begin{align}\label{3d32}
	E^{(2)}_\mu(n,T) & = g_0 - \frac{1}{2\tilde g_2 \gamma_\tau} \lp \frac{\tilde\mu}{W(0)}\rp  + n_c \bar M - \frac{1}{2} n_c^2 \tilde D(0) - \frac{1}{24} \frac{g_3^4}{g_4^3}.
\end{align}
Plot of the pressure as a function of density at $T=T_c$ is represented in Figure~\ref{fig5}. It shows, that $n_g$ is the critical density value. 

\section{Phase behavior of a cell fluid model}

In Sections 3 and 4, the equation $ P = P (\tau, M) $ is obtained for the temperature range above and below the critical one. Section 5 states that the range of the chemical potential $ M $ is limited by two characteristic curves $ M^* $ \eqref{3d17} and $ M^{**} $ \eqref{3d19}, respectively,
\begin{equation}\label{5d1}
	M^* \leq M \leq M^{**}.
\end{equation}
Figure~\ref{fig2} shows the 3D plot of the equation of state $ P (\tau, M) $ for the temperature domain $ T <T_c $. It has a fracture along with the line $ M = 0 $, which indicates a first-order phase transition, since the first derivative of pressure with respect to the chemical potential $ M $ on the line $ M = 0 $ will be different from the left-hand side and the right-hand side. In addition, Figure~\ref{fig4} shows the line $ M_q (\tau) $, which delimits regions of the surface $ P (\tau, M) $, in which there are either one or three real solutions of the equation \eqref{2d16} in the unknown $ \bar \rho_0 $. Finding $ \bar \rho_0 $ makes it possible to calculate the grand partition function of the cell fluid model and establish the existence of a first-order phase transition in it. Figure~\ref{fig6} illustrates the coexistence curve and the spinodal of the present model. Each of these curves ends at temperature $ T = T^* <T_c $.

\section{Conclusions}

The results of the present research state that the cell fluid model is suitable for the description of a first-order phase transition. The interaction potential of such a model, in addition to the attractive and repulsive interaction (such as the Morse potential), should include additional repulsive interaction (a reference system). The use of supplementary interaction enables studying a phase behavior of the model in a wide range of density and temperature.

A feature of the proposed approach is the use of the grand canonical ensemble, in which temperature and chemical potential are independent variables. First, the equation of state of the cell fluid model is obtained in terms of these variables and then transposed to the temperature-density plane. For this purpose, we derived and solved the equation, which relates the average number of particles (density) to the chemical potential. The solutions are functions of temperature. Such dependence is found to be substantially nonlinear. Moreover, it implies a limited lower and upper range of density and chemical potential. We prove that the minimum and maximum admissible values of the chemical potential approach each other below the critical temperature. An estimated value of temperature $ T^* $, at which they coincide, is $ T^* \approx 0.38 T_c $ (for the sodium-specific interaction potential parameters given in~\cite{ref9}). The conclusion on the existence of a first-order phase transition in the temperature range $ T^* <T <T_c $, as well as the absence of the liquid-gas phase transition in the temperature range $ T \leq T^* $ is obvious. The temperature of $ T^* $ can be interpreted as the triple point temperature. We obtain both the maximum density value $ n_{max} $ for a liquid phase and the minimum density value in a gaseous phase $ n_{min} $ (formula \eqref{3d20}). Each of these densities corresponds to the temperature specified in the formula \eqref{3d18}. 

The proposed method for calculating phase behavior is possible to extend to other systems with known parameters of interaction potentials. 

The equation obtained in this paper is not suitable near the critical point ($ \tau <\tau^* \approx 10^{-2} $). In this case, one should count the effects of long-range fluctuations since they play a major role in the description of the critical region above and below the critical point. In this regard, there are theoretical approaches \cite{ref14,ref17,ref18,ref19} and experiments \cite{ref20,ref21,ref22} confirming the presence of non-classical critical exponents of compressibility, heat capacity, correlation length, etc. We have also taken into account fluctuations (within the approach represented in the present paper) based on the calculation of the grand thermodynamic potential, considering the contributions of the collective variables $ \rho_{\vk} $ with $ | \vk | \neq 0 $ in \cite{ref6}.

\end{document}